# Spatial periodic and homogeneous transverse stress loading on ITER TF Nb$_3$Sn bronze and internal tin strand

A. Nijhuis, R.P. Pompe van Meerdervoort, W.A.J. Wessel

*Abstract*— The transport properties of the superconducting Nb$_3$Sn layers in the strands strongly depend on the strain state. Knowledge of the influence of axial strain, periodic bending and contact stress on the critical current ($I_c$) of the used Nb$_3$Sn strands is inevitable to gain sufficient confidence in an economic design and stable operation of ITER CICCs. In the past years we have measured the Ic and n-value of various ITER Nb$_3$Sn strands with different layout in the TARSIS facility, when subjected to spatial periodic contact stress at a temperature of 4.2 K and in a magnet field of 12 T. Recently we have made the setup suitable for application of homogeneous load along the length of the wire (125 mm) in order to evaluate possible differences related to spatial stress and possible current distribution. We present an overview of the results obtained so far on an ITER TF bronze and internal tin strand.

## I. INTRODUCTION

THE envisaged CICCs for ITER consist out of more than 1000 strands with a strand diameter of about 0.8 mm. The conductors are cabled by twisting in several stages following a wavy pattern. During magnet operation, they carry currents in the order of 50-70 kA in a magnetic field locally exceeding 12 T, hence subjecting them to severe transverse loading (bending and contact stress) due to the huge Lorentz forces. Bending and pinching at the contacts of Nb$_3$Sn strand affects the critical current and creates a periodic strain variation along the filaments and a strain distribution along the wires [1]. The magnitude and periodicity of the bending and contact strain (distribution) in combination with the strand stiffness and the ability of a strand to redistribute the current between the filaments, determines the impact on the critical current and n-value of the final cabled conductor.

The single strand properties in terms of the composite stiffness and the change of the Ic and n-value under various transverse load conditions are crucial input parameters for various models. For this reason, the TARSIS setup with different probes to simulate equivalently ITER magnet operating conditions on single strands was build. Probes were developed for periodical strand bending with cyclic loading using different bending wavelengths, for the characterisation of the axial tensile stress strain behaviour of strands and sub-size cables (several references can be found in [1, 2]).

The impact of local transverse loads at strand crossings has not received much attention worldwide since the first papers appeared on this subject [3-5], although for accelerator type of conductors additional effort was done [6, 7]. Recent work on homogenous stress and Nb$_3$Sn can be found in [8]. Transverse stresses can be quite severe at one side of particularly large ITER type of cables, depending on the cabling pattern, resulting in to an accumulated load that can reach peak values in the order of 100 MPa. In particular strand types that are not well mechanically protected against severe local stresses, for example strands with large Kirkendall voids (due to the reaction of tin with niobium leaving voids where the tin was located before reaction heat treatment), may be more susceptible to this type of load [6]. The aim of this work is to quantify the influence of the transverse stress and in particular investigate the influence of stress distribution in terms of periodicity. This may lead to a better understanding of the role of the interfilament current redistribution on the average strand electric field and its distribution [9, 10].

Here, we describe the modified probe to study the difference of transverse contact load on perpendicularly crossing and pinching strands and we present the first results on a bronze (BR) and internal tin (IT) ITER Nb$_3$Sn strand.

## II. EXPERIMENTAL SET UP

The design requirements of the TARSIS setup and its probe for periodic contact stress is described in detail in [2]. The advantage of the TARSIS setup is that besides precise measurement of the strand amplitude of deflection or deformation, also the applied force is monitored for transverse stress-strain analysis. Photographs of the probe with wire sample, crossing strands and homogeneous load cap is shown in Fig. 1. The wire runs along the circumference of a cylindrical plate with an outer diameter of 47 mm and is entirely supported by the flange of the outer barrel. The cylindrical support plate has radial grooves on top from centre to the outer perimeter to guide the crossing strand sections. A cap having a flat bottom is placed on top, transferring the load from the crossing strands between cap and support plate with grooves to the circular sample wire. The load on the crossing strands is utilised by moving a thrust bar in the direction of the

Manuscript received 12 September 2011. This work was supported in part by the ITER International Organisation, Cadarache (France) and F4E Barcelona (Spain).

A. Nijhuis, R.P. Pompe van Meerdervoort, W.A.J. Wessel are with the University of Twente, Faculty of Science & Technology, Energy, Materials & Systems, 7500AE Enschede, The Netherlands; e-mail: a.nijhuis@utwente.nl.

sample. The force applied with the thrust bar is measured with a load cell outside the cryostat. For the actual deflection measurement, we attach an extensometer in each of the three slots of the barrel. The current sharing voltage per unit length (*E*) is measured by means of three pairs of voltage taps corresponding with the location of the three extensometers, so the homogeneity of the load is probed as well. The spatial periodicity (wavelength) for crossing strands is 4.7 mm and the wire length subjected to homogeneous load is 125 mm.

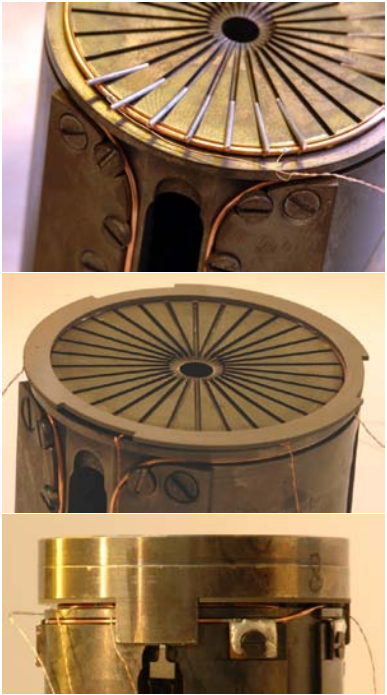

Fig. 1 A photograph of the of the X-strands probe (top) with test wire (soldered voltage taps are visible) and crossing wires. In the middle the flat ring is shown for homogeneous load and below with the top plate mounted.

### III. STRAND AND SAMPLE PREPARATION

The sample preparation is described in detail in [2]. The two $Nb_3Sn$ samples for the tests, one bronze and one internal tin type, are taken from the ITER TF production. The diameter is 0.82 mm. The wire is wound on the cylinder prior to the heat treatment with an axial pre-stress of 25 MPa. The sample is mounted on the sample holder under a slight transverse pre-load before heat treatment to guarantee contact with the load points and avoid settling alterations during the initial loads of the test. The applied heat treatment schedules are as specified by ITER IO. The distance between the voltage taps is 23 mm for the X-strands probe and 20 mm in the case of the homogenous stress arrangement. The filament twist pitch of the strand is about 15 mm.

### IV. EXPERIMENTAL RESULTS

#### A. Critical current and n-value

The measurements are performed at a temperature of 4.2 K and a magnetic field of 12 T. The voltage current traces (V-I) at various applied contact loads are recorded from 1 µV/m up to an electric field level of at least 100 µV/m. Increasing inter-filament current transfer results in a lower n-value. The longitudinal electric field in a strand $E_z$ is a non-linear function of the current, of the temperature and of the magnetic field. It is calculated by solving the following equations where $A_{sc}$ and $A_m$ are the area of the cross section which are occupied by superconductor and matrix material respectively, $J_{sc}$ and $J_m$ are the current density in the superconductor and in the matrix material, $E_c$ and $J_c$ are the critical parameters of the superconductors and $\rho_m$ is the electrical resistivity of the matrix material. The axial strain of the $Nb_3Sn$ filaments and the stabiliser is represented by ε.

$$I = J_{sc}A_{sc} + J_m A_m \quad ; \quad E = E_c\left(\frac{J_{sc}}{J_c(T,B,\varepsilon)}\right)^{n(T,B,\varepsilon)} = \rho_m(T,B,\varepsilon)J_m \quad (1)$$

For a given current and a possible change in the matrix resistivity, the $I_c$ will change accordingly.

The *n*-values are normally determined from the V-I curves in the electric field range of 10-100 µV/m with the commonly used power law V-I representation from relation (1). The V-I characteristic, as the current increases, gradually travels from no to full current transfer and the characteristic strongly depends on the local strain distribution and interfilament resistivity. Classically we assume strands to have a constant n value obeying a power law fit but for increasing strain, the V-I curve does not follow anymore the power law. This is illustrated by a few selected V-I curves at different loads in Fig. 2 where it clear that not only the steepness of the curves changes with increasing load but also the shape.

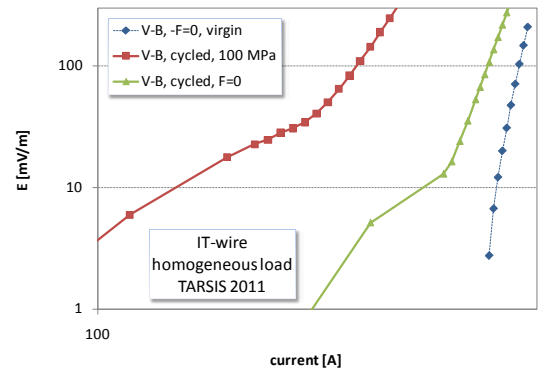

Fig. 2 Examples of measured voltage current curves for virgin state, a stress level of 100 MPa and release to zero after 100 MPa.

Nevertheless, we keep using the classical formulation of the n-value, reflecting the steepness of the V-I transition between 10 and 100 µV/m, in order to avoid confusion in discussions with reference to the traditional definition.

The critical current versus the applied load for a periodicity of 4.7 mm of crossing strands is presented in Fig. 3. The critical currents under load, taken at an electric field of 10 µV/m, are normalized to the $I_{c0}$ in the virgin (unloaded) condition of the strand for display in the plots. The load is increased stepwise with intermediate release to zero in order to evaluate the irreversibility stress level. The irreversibility threshold may or may not be related to damage in the $Nb_3Sn$ layers but is also associated with plastic strand deformation. The applied stress (MPa) is calculated as the projected cross sectional area. So for the crossing strands (X-strand) the area is taken as the force per strand crossing, divided by the square of the strand diameter. For homogeneous load the stress is taken as the force divided by the (strand diameter times the loaded length).

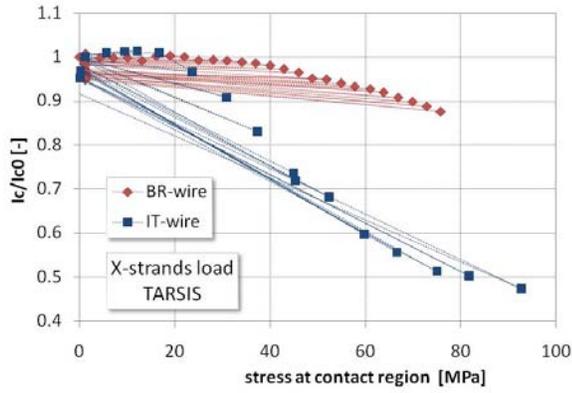

Fig. 3 The reduction in critical current for the BR and IT strand versus the applied stress for the X-strands test.

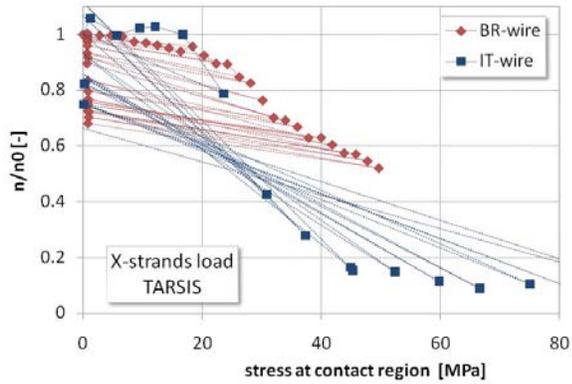

Fig. 4 The reduction of the n-value for the BR and IT strand versus the applied stress for for the X-strands test.

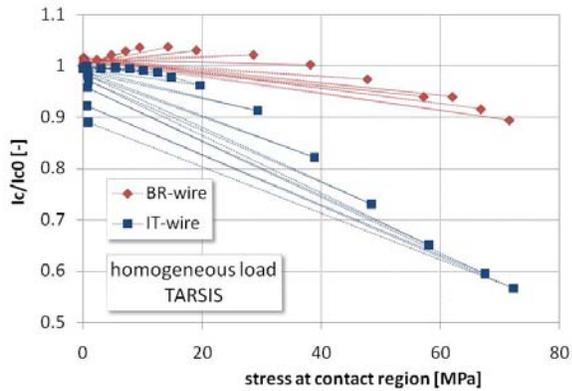

Fig. 5 The reduction in critical current for the BR and IT strand versus the applied stress for homogeneous distributed load.

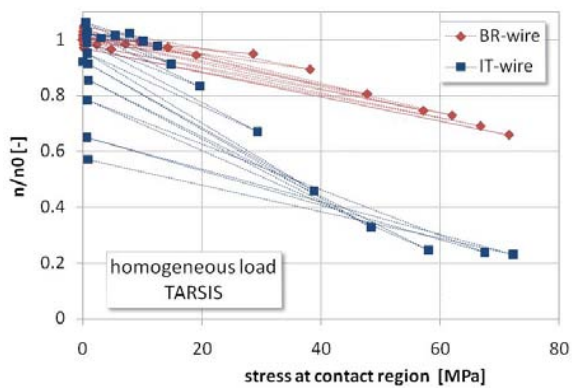

Fig. 6 The reduction of the n-value for the BR and IT strand versus the applied stress for homogeneous distributed load.

The evolution of the n-value with applied stress is depicted in Fig. 4, the critical current and n-value versus the applied load for homogeneous load is presented in Fig. 5 and Fig. 6.

The BR-wire is clearly less sensitive then the IT-wire, to both transverse X-strand and homogeneous stress at higher load levels. However, for X-strands the irreversibility for BR-wire starts to occur at about 20 MPa while for IT-wire this is much higher (about 40 MPa). For homogeneous load we find an increase of the Ic with applied load for the BR-wire up to 30 MPa, likely attributed to stress release in the strand matrix, while the IT-wire shows only a gradual decrease of the $I_c$. The irreversibility stress level for homogeneous load (likely best represented by an irreversible degradation of the $n$-value, is found at 60 MPa for BR-wire while for the IT-wire it is at about 20 MPa.

In the previous figures, the reduction of the $I_c$ and $n$-value is plot against the applied stress in the contact zone. When we plot the reduction of the $I_c$ against the length of the strand which is subjected to the load, then the sensitivity is much higher for X-strands as the load is concentrated in the contact points giving high local stress (see Fig. 7, we have observed a similar tendency for BR). For homogeneous load the applied force is distributed along the length and its impact is obviously much less. Obviously this reflects the different condition in a CICC with long or short twist pitches [1]. When we compare the influence of local periodic and homogeneous stress, the reduction in $I_c$ follows exactly the same tendency for both stress distributions (see Fig. 8). Similar tendencies are observed for BR wire (Fig. 9) and the evolution of the $n$-value (Fig. 10 and Fig. 11).

## V.  DISCUSSION

This means that the absolute value of the stress determines the reduction in Ic and n-value and the spatial distribution seems not relevant for the measured conditions; X-strand periodicity of 4.7 mm and homogeneous distribution. In fact this is well in agreement with the results found for periodic bending with different bending wavelengths. For spatial periodic bending of three $Nb_3Sn$ strands, manufactured with different processing techniques, we found that the critical current reduction versus applied bending strain was similar for all applied wavelengths of 5.0, 7.2, 8.3 and 10.0 mm [11, 12]. Also here the periodicity is not important but only the applied peak bending strain.

The next step in this study is to test a spatial periodicity with larger spacing between the crossing strands and to evaluate the required length for interfilament current redistribution in order to reach lower average electric fields along the wires. This will be evaluated with the strand model presently under development and the ongoing studies on intrastrand resistance distribution and current transfer length [9, 10] and additional voltage taps. The interstrand current transfer length in CICCs depends on the loading history and void fraction but is expected to go with relatively long interstrand current transfer lengths.

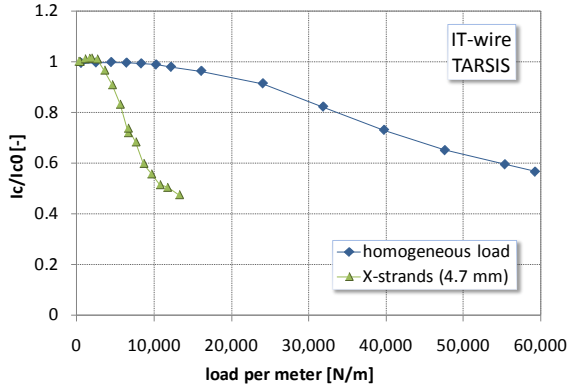

Fig. 7 The reduction of the Ic versus the applied load per meter strand (IT) for X-strands and homogeneous load.

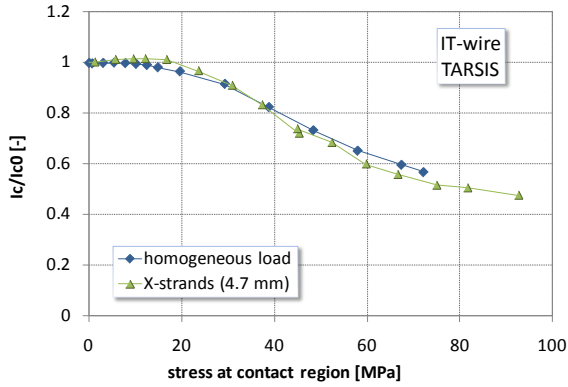

Fig. 8 The reduction of the Ic versus the applied stress (IT) for X-strands and homogeneous load.

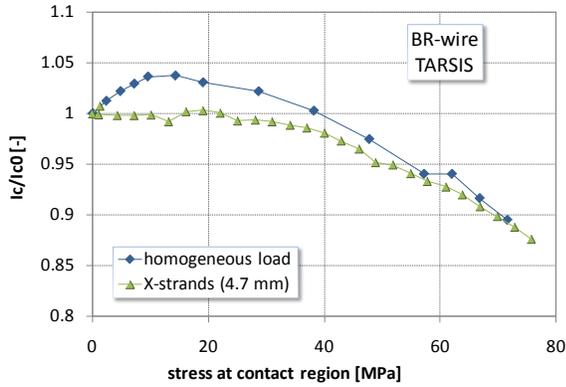

Fig. 9 The reduction of the Ic versus the applied stress (BR) for X-strands and homogeneous load.

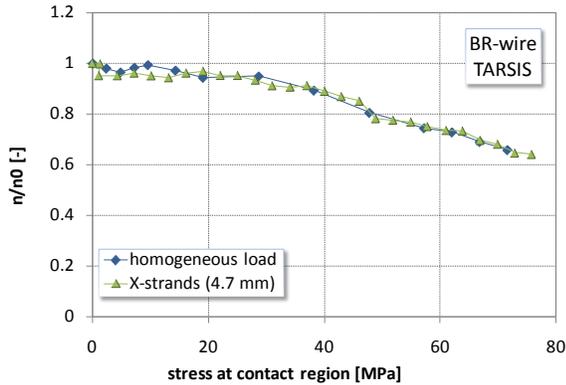

Fig. 10 The reduction of the n-value versus the applied stress (BR) for X-strands and homogeneous load.

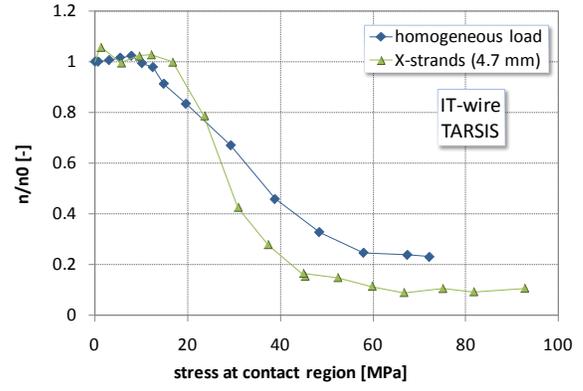

Fig. 11 The reduction of the n-value versus the applied stress (IT) for X-strands and homogeneous load.

## VI. CONCLUSIONS

The effect of spatial periodic loading with crossing strands and homogeneously distributed load on the transport properties is compared for a bronze and internal tin ITER TF $Nb_3Sn$ strand.

The irreversible stress level differs for both load distributions and strand type, varying between 20 and 60 MPa. The absolute value of the stress determines the reduction in $I_c$ and $n$-value and the spatial distribution seems not relevant for a crossing strand periodicity of 4.7 mm compared to homogeneous distribution. This is well in agreement with the results found for periodic bending with different bending wavelengths. More work is needed to explore the relation between periodicity and performance reduction.